\documentclass[runningheads]{svmult}

\usepackage{makeidx}   % allows index generation
\usepackage{graphicx}  % standard LaTeX graphics tool
\usepackage{subeqnar}  % subnumbers individual equations
\usepackage{multicol}  % used for the two-column index
\usepackage{physprbb}  % modified textarea for proceedings,
\makeindex             % used for the subject index
\begin{document}
\title*{New Evolutionary Synthesis Tool for Modelling Young Star Clusters in Merging Galaxies}
\toctitle{New Evolutionary Synthesis Tool for Modelling Young Star Clusters in Merging Galaxies}
% allows explicit linebreak for the table of content
%
%
\titlerunning{New ES Tool for Modelling YSCs in Merging Galaxies}
% allows abbreviation of title, if the full title is too long
% to fit in the running head
%
\author{Peter Anders\inst{1}
\and Richard de Grijs\inst{2}
\and Uta Fritze - v. Alvensleben\inst{1}}

\authorrunning{Peter Anders et al.}
% if there are more than two authors,
% please abbreviate author list for running head
%
%
\institute{Universit\"ats-Sternwarte G\"ottingen, Geismarlandstr. 11, 37083
G\"ottingen, Germany
\and Institute of Astronomy, Madingley Road, Cambridge CB3 0HA, UK}

\maketitle              % typesets the title of the contribution

\section{Introduction}
Globular cluster systems (GCSs) are vital tools for investigating the violent star formation histories of their host galaxies. This violence could e.g. have been triggered by galaxy interactions or mergers. The basic observational properties of a GCS are its luminosity function and color distributions (number of clusters per luminosity resp. color bin).

A large number of observed GCSs show bimodal color distributions, which, by comparison with evolutionary synthesis (ES) models, can be translated into bimodality in metallicity and/or age. An additional uncertainty comes into play when one considers extinction within the host galaxy.

These effects can be disentangled either by obtaining spectroscopic data for the clusters or by imaging observations in at least four passbands. This allows us then to discriminate between various formation scenarios of GCSs, e.g. the merger scenario by Ashman \& Zepf \cite{AZ} and the multi-phase collapse model by Forbes et. al. \cite{Forbes}.

Young and metal-rich star cluster populations are seen to form in interacting and merging galaxies. We analyse multi-wavelength broad-band observations of these young cluster systems provided by the ASTROVIRTEL project.

\section{Modelling multi-wavelength star cluster data}

\begin{figure}[ht]
\includegraphics[width=0.5\columnwidth]{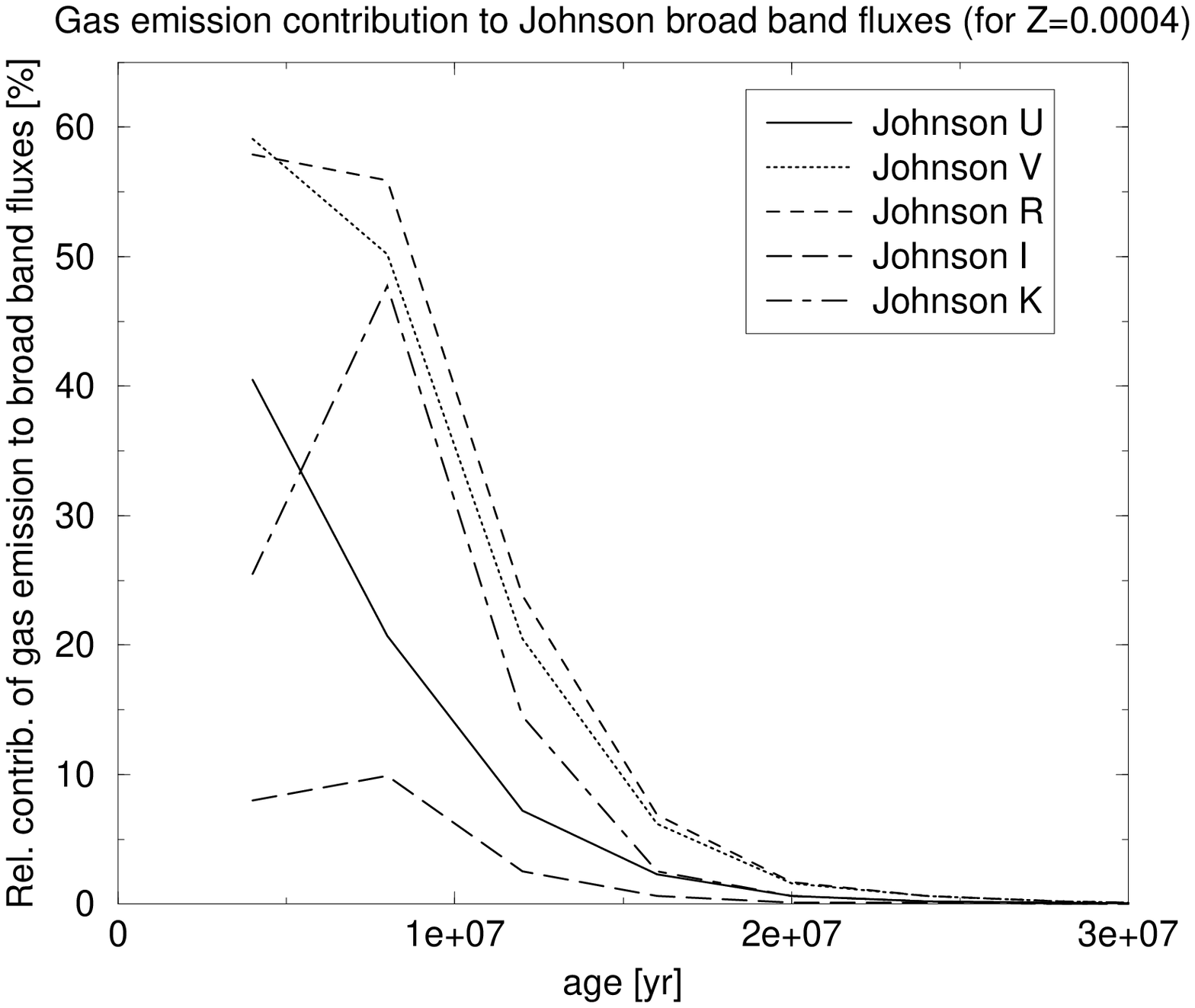}
\includegraphics[width=0.5\columnwidth]{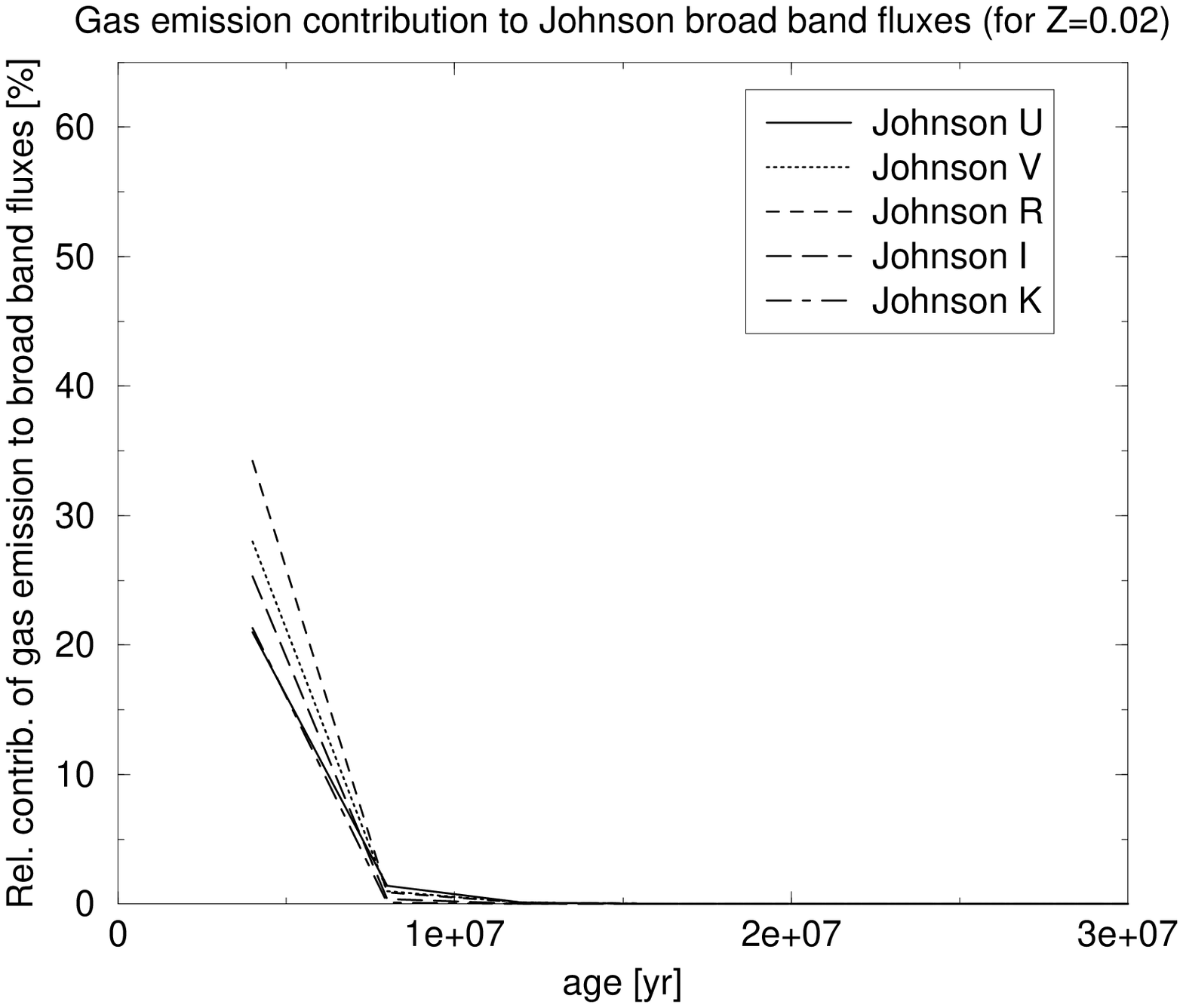}
\caption{Time evolution of the gaseous emission contribution to broad-band fluxes in Johnson passbands $U, B, V, I$, and $K$ at low metallicity ${\rm Z = 0.0004}$ (left panel) and solar metallicity (right panel).}
\label{fig1}
\end{figure}

We have further extended the G\"ottingen evolutionary synthesis code by including the effects of gaseous emission. The gaseous emission contributes significantly to the integrated light of stellar populations younger than $3 \cdot 10^7$ years \cite{Andersa}; see Figure \ref{fig1}. In addition, the effect of various amounts of {\sl internal} dust extinction has been included. 

\begin{figure}[!ht]
\includegraphics[width=0.5\columnwidth]{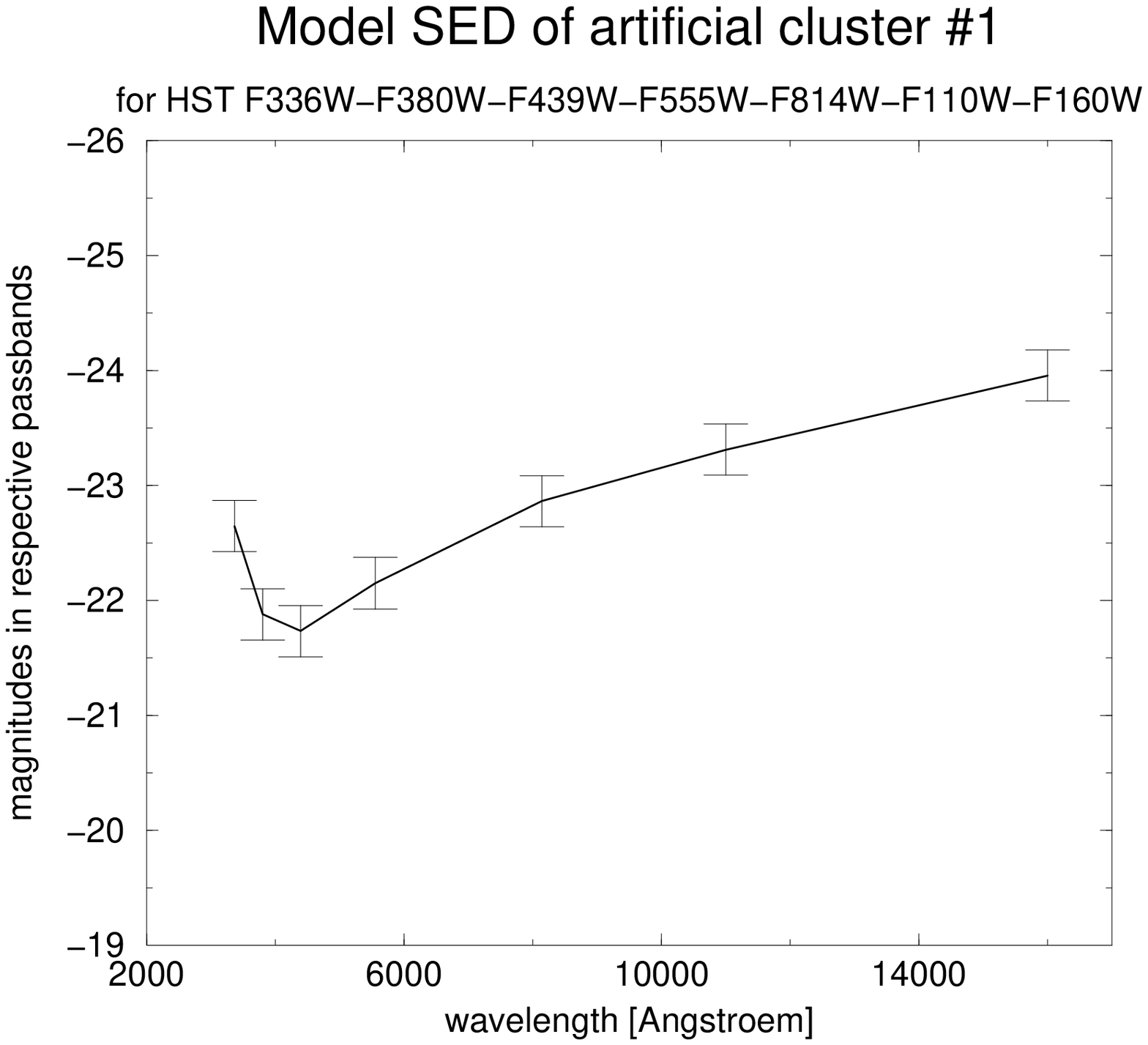}
\includegraphics[width=0.5\columnwidth]{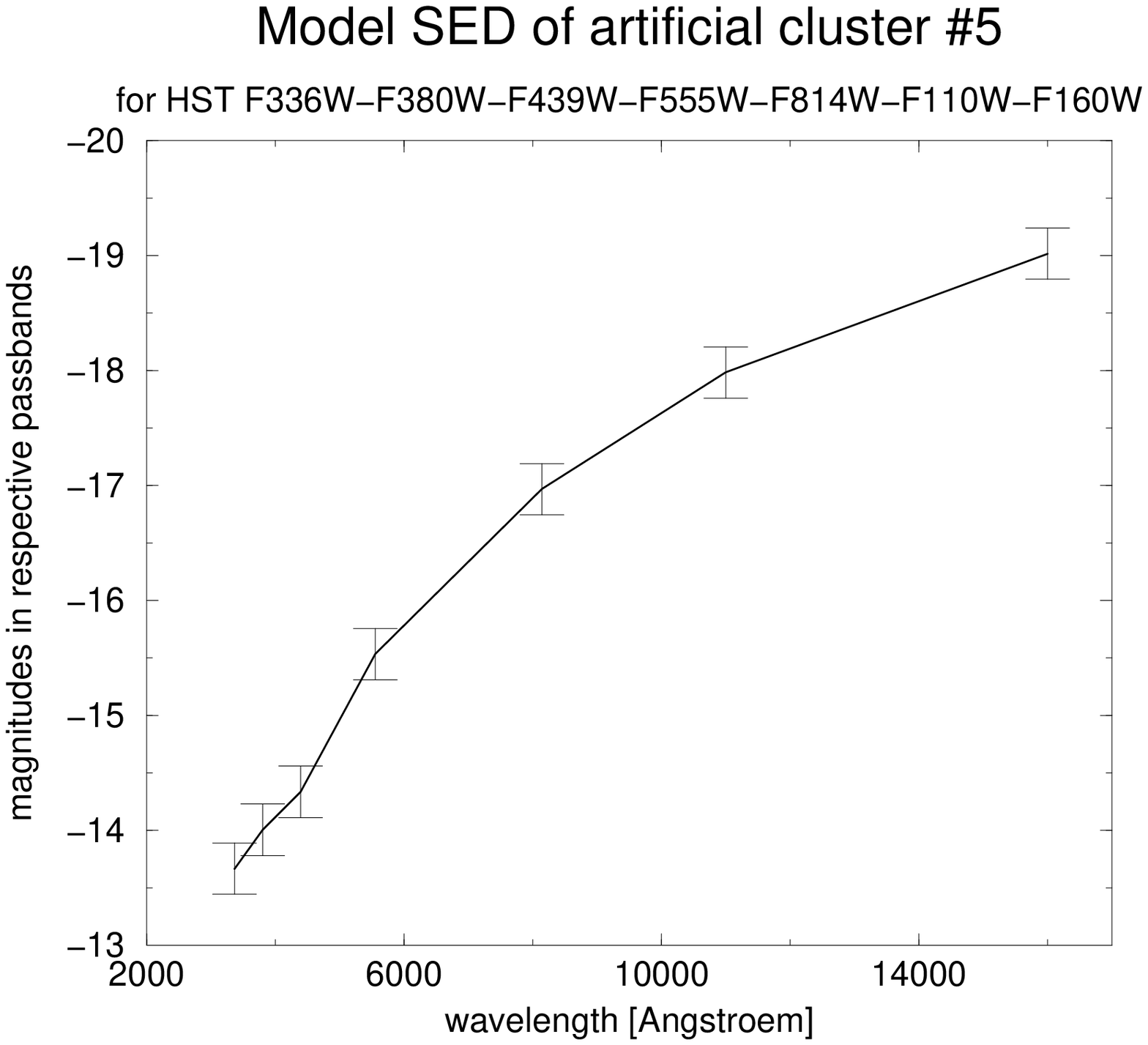}
\caption{Model SEDs for 2 artificial clusters \#1 (left panel) and \#5 (right panel) with solar metallicity, E(B-V)=0.1 and ages 8 Myr and 10 Gyr respectively.}
\label{fig2}
\end{figure}

The simultaneous determination of a cluster's age, metallicity, extinction and mass is achieved by comparing an appropriate grid of ES models with the observed spectral energy distribution (SED) in a least-squares sense. Examples for model SEDs are shown in Figure \ref{fig2}.

Due to the well-known age-metallicity degeneracy (and a similar age-extinction degeneracy) at optical wavelengths (and the scaling of a cluster's luminosities with its mass), the use of multi-passband observations is essential to determine these parameters independently. 

\subsection{Results for Artificial Clusters}

We constructed five artificial clusters with known properties, i.e., solar metallicity, low extinction $E(B-V)=0.1$, ages of 8, 60, and 200 Myr, and 1 and 10 Gyr. The SEDs for these artificial clusters were taken from our model grid, and a typical observational error of 0.1 mag was assumed. Subsequently, we redetermined the best-fit cluster parameters using our least-squares fitting algorithm, and compared the results to the original values. We constructed a number of additional clusters characterized by color offsets (too red or too blue by 0.1 mag) to assess the accuracy of our model fits. From this study we conclude that:

\begin{enumerate}

\item The choice of passband combination required for a proper determination of the basic cluster parameters (age, metallicity, extinction) can be optimized towards the expected mean parameter values.

\item In order to determine reliable ages, the use of ultraviolet (UV) passbands is vital for all ages. Including a near-infrared (NIR) passband improves the accuracy further. 

\item For metallicity determinations, NIR observations are most important, while the UV is essential for young systems (ages $<$ 100 Myr), due to the metallicity-dependence of the emission line strengths from non-hydrogen elements.

\item The extinction is best determined using UV + optical colors.

\end{enumerate}

In general, the accuracy and the best passband combination are a strong function of the age of the cluster's stellar population \cite{Andersb}. The availability of observations spanning the entire wavelength range, from UV to NIR, allows to constrain all individual parameters (age, metallicity, internal extinction and mass) most efficiently. 

\section{Application and first results: NGC 1569}

The irregular dwarf galaxy NGC 1569 is commonly classified as a (post) starburst galaxy with H{\sc ii} region metallicity $\sim$ (0.2 -- 0.5) $Z_\odot$. High-resolution multi-passband imaging data were provided by the ESO / ST-ECF ASTROVIRTEL project ``The Evolution and Environmental Dependence of Star Cluster Luminosity Functions'' (PI R. de Grijs). In addition to the two well-studied super star clusters in its main disk, they reveal a number of fainter objects resembling compact star clusters (see also \cite{Hunter}).

\begin{figure}[!ht]
\includegraphics[width=0.5\columnwidth]{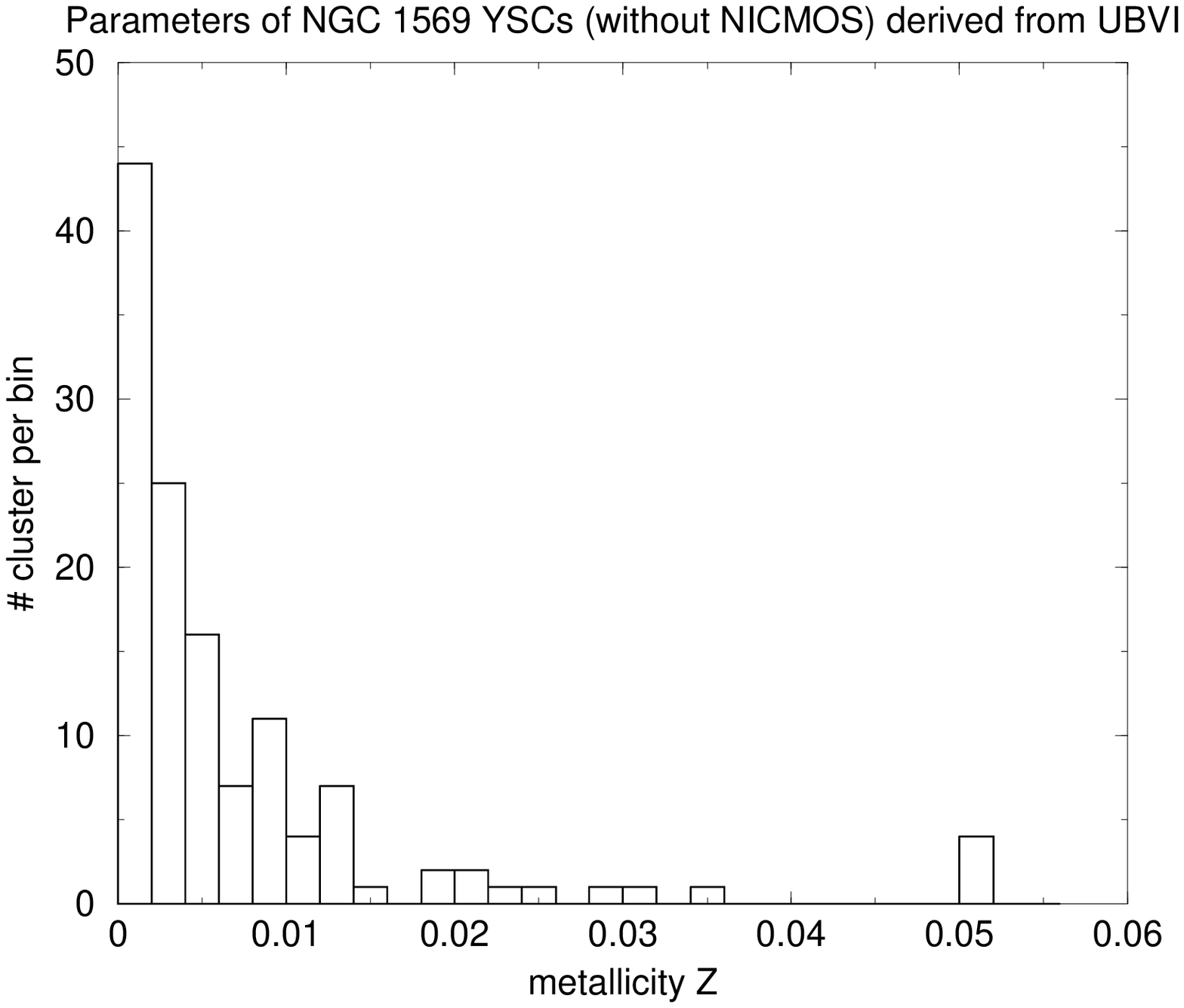}
\includegraphics[width=0.5\columnwidth]{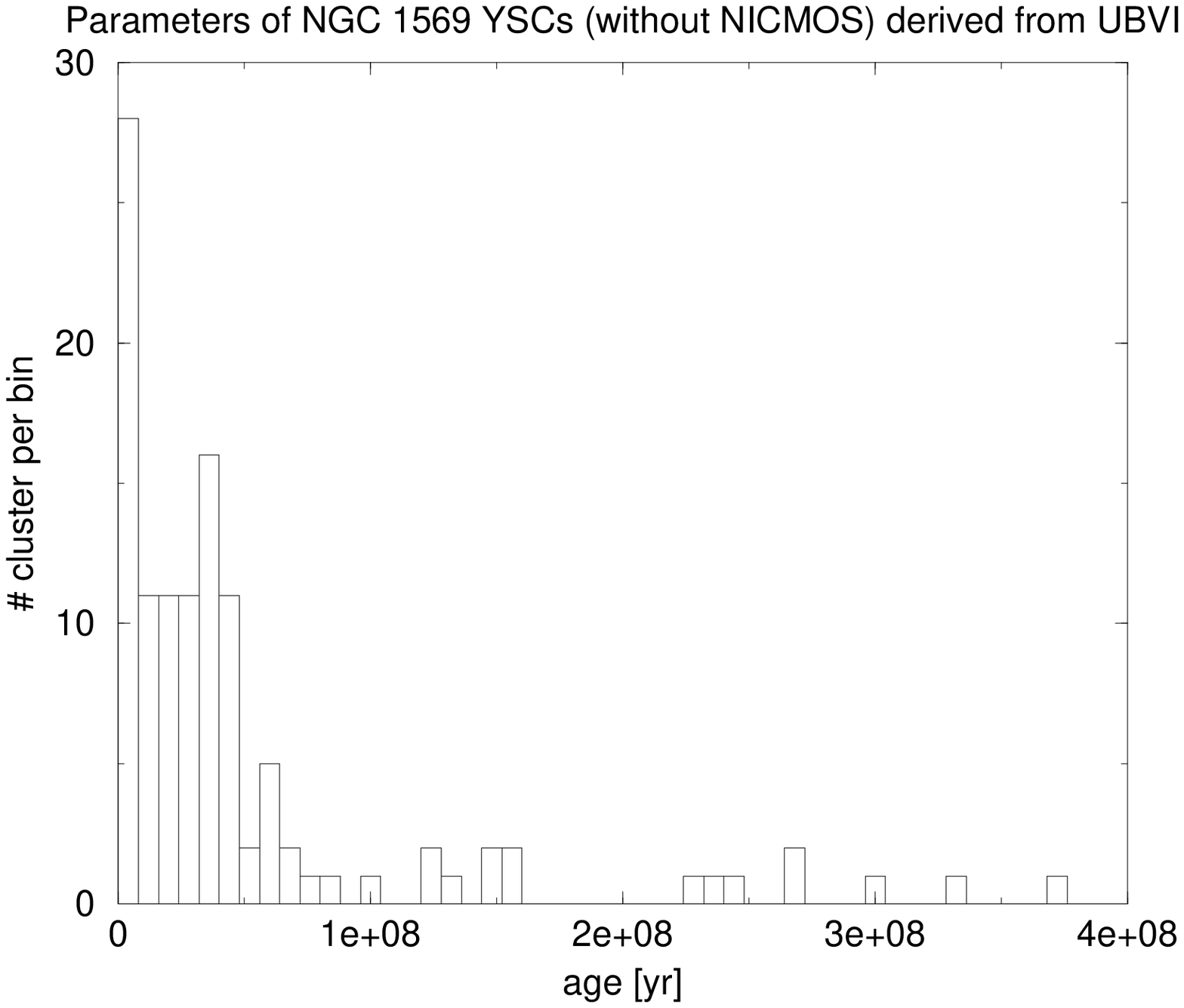}
\caption{Left panel: Metallicity distribution of young star clusters in NGC 1569, derived from $U, B, V$ and $I$; right panel: corresponding age distribution (during the burst; older clusters, up to 5.2 Gyr old, are also present).}
\label{fig3}
\end{figure}

Our fitting algorithm provides evidence for continuous star formation starting at least 5.2 Gyr ago. The recent burst, however, started around 80 Myr ago, with a possible peak of cluster formation around 35 Myr ago and a major peak in the cluster formation rate in the youngest age bin $\le$ 8 Myr (see Figure \ref{fig3}). These results are consistent with previous results \cite{Aloisi,Greggio} regarding the bulk of the starburst activity in general and the formation of the two super star clusters in particular. As seen in previous studies, the most recent enhanced epoch of star formation coincides well with the formation of morphologically peculiar features, like $H_\alpha$ filaments and ``superbubbles'' \cite{Waller}. 

An accompanying H{\sc i} cloud is located at 5 kpc from NGC 1569, connected to it by an H{\sc i} bridge, suggesting a tidal interaction as a possible cause for the starburst \cite{StilIsrael}.

We determine masses for the individual star clusters by comparing the observed luminosities with the luminosities of our model cluster of given mass at each individual's cluster's age, metallicity and extinction. We find masses up to $10^6 M_\odot$. The metallicity distribution is broader than found in previous surveys, which mainly focussed on the ISM \cite{Devost,Kobulnicky}. This may imply different chemical enrichment processes in the star clusters vs. the ISM, or selection effects caused by the different spatial coverage. The internal dust extinction is found to be low ($E(B-V) < 0.35$, with the vast majority of the clusters being affected by $E(B-V) < 0.1$ only) \cite{Andersc}.

\section{Acknowledgements}
Very efficient support given by ASTROVIRTEL, a Project funded by the European Commission under FP5 Contract No. HPRI-CT-1999-00081, is gratefully acknowledged. P. Anders is partially supported by DFG grant Fr 916/11.

\end{document}